\documentclass[10pt,journal,compsoc]{IEEEtran}

\usepackage{multirow}
\usepackage{comment}
\usepackage{adjustbox}
\usepackage{graphicx}
\usepackage{textcomp}
\usepackage{gensymb}
\usepackage[caption=false,font=footnotesize]{subfig}
\usepackage{float}
\usepackage{xcolor}
\usepackage[utf8]{inputenc}
\usepackage{soul}

\ifCLASSINFOpdf
\else

\fi

\newcommand{\jg}[1]{{\textcolor{red}{JG: #1}}}
\newcommand{\pab}[1]{{\textcolor{magenta}{[PP] #1}}}
\newcommand{\moc}{}

\usepackage{url}
\usepackage{array}

\begin{document}

\title{Methodology to Assess Quality, Presence, Empathy, Attitude, and Attention in \moc{
360-degree Videos for Immersive Communications}
}

\author{Marta~Orduna, Pablo~P\'erez, Jes\'us~Guti\'errez,
        and~Narciso~Garc\'ia 
\IEEEcompsocitemizethanks{\IEEEcompsocthanksitem M. Orduna, J. Guti\'errez and N. Garc\'ia are with the Grupo de Tratamiento de Imágenes, Information Processing and Telecommunications Center and Escuela Técnica Superior de Ingenieros de Telecomunicación, Universidad Politécnica de Madrid, Madrid 28040, Spain. E-mail: \{moc, jgs, narciso\}@gti.ssr.upm.es.\protect \\
\IEEEcompsocthanksitem P. P\'erez is with  Application Platforms and Software Systems Labs, Nokia Bell Labs, Madrid 28050, Spain. E-mail: pablo.perez@nokia-bell-labs.com}
\thanks{Manuscript received December 4, 2020; revised October 1, 2021.}}

\markboth{IEEE Transactions on Affective Computing ~Vol.~XX, No.~X, XXX~XXXX}%
{Shell \MakeLowercase{\textit{et al.}}: Bare Demo of IEEEtran.cls for Computer Society Journals}

\IEEEtitleabstractindextext{%
\begin{abstract}
This paper \moc{proposes} a methodology to assess video quality, spatial and social presence, empathy, attitude, and attention in 360-degree videos \moc{for immersive communications}. \moc{The methodology is validated in an experiment which simulates} an immersive communication environment where participants attend three conversations of different genre (everyday conversation, educational, and discussion) and from actor and observer acquisition perspectives. \moc{We consider three experimental conditions:} (A) visualizing and rating the perceptual quality of contents in a Head-Mounted Display~(HMD), (B) visualizing the contents in an HMD, and (C) visualizing the contents in an HMD where participants can see their hands and take notes. In all conditions participants visualize the same 360-degree videos, designed and acquired in the context of international experiences. Fifty-four participants were evenly distributed among A, B, and C conditions taking into account their international experience backgrounds (working or studying in a foreign country), obtaining a balanced and diverse sample of participants. In this paper, video quality is evaluated with Single-Stimulus Discrete Quality Evaluation~(SSDQE) methodology. Spatial and social presence are evaluated with questionnaires adapted from the literature. Initial empathy is assessed with Interpersonal Reactivity Index (IRI) and a questionnaire is designed to evaluate the attitude \moc{after the visualization of each video}. Attention is evaluated with three questions about the conversations of the contents that had pass/fail answers. The results from the subjective test validate the proposed methodology in \moc{immersive} communications, showing that video quality experiments can be adapted to conditions imposed by experiments focused on the evaluation of socioemotional features in terms of contents of long-duration, \moc{different} acquisition perspectives, and genre. In addition, the positive results related to the sense of social and spatial presence imply that technology can be relevant in the analyzed use case. Other main result is that the acquisition perspective greatly influences social presence. 
 Finally, the annotated dataset, Student Experiences Around the World dataset~(SEAW-dataset), obtained from the experiment is made publicly available for the research community.
\end{abstract}

}

\maketitle

\IEEEdisplaynontitleabstractindextext

%
\IEEEpeerreviewmaketitle

\section{Introduction}\label{sec:introduction}

Virtual Reality~(VR) is an emerging field that is achieving great interest in applications for social purposes, such as assistive, entertainment or educational applications, and also in teleconferencing scenarios~\cite{gunkel2018virtual, li2020social}.

  \begin{figure}[t!]
  \includegraphics[width=1\columnwidth]{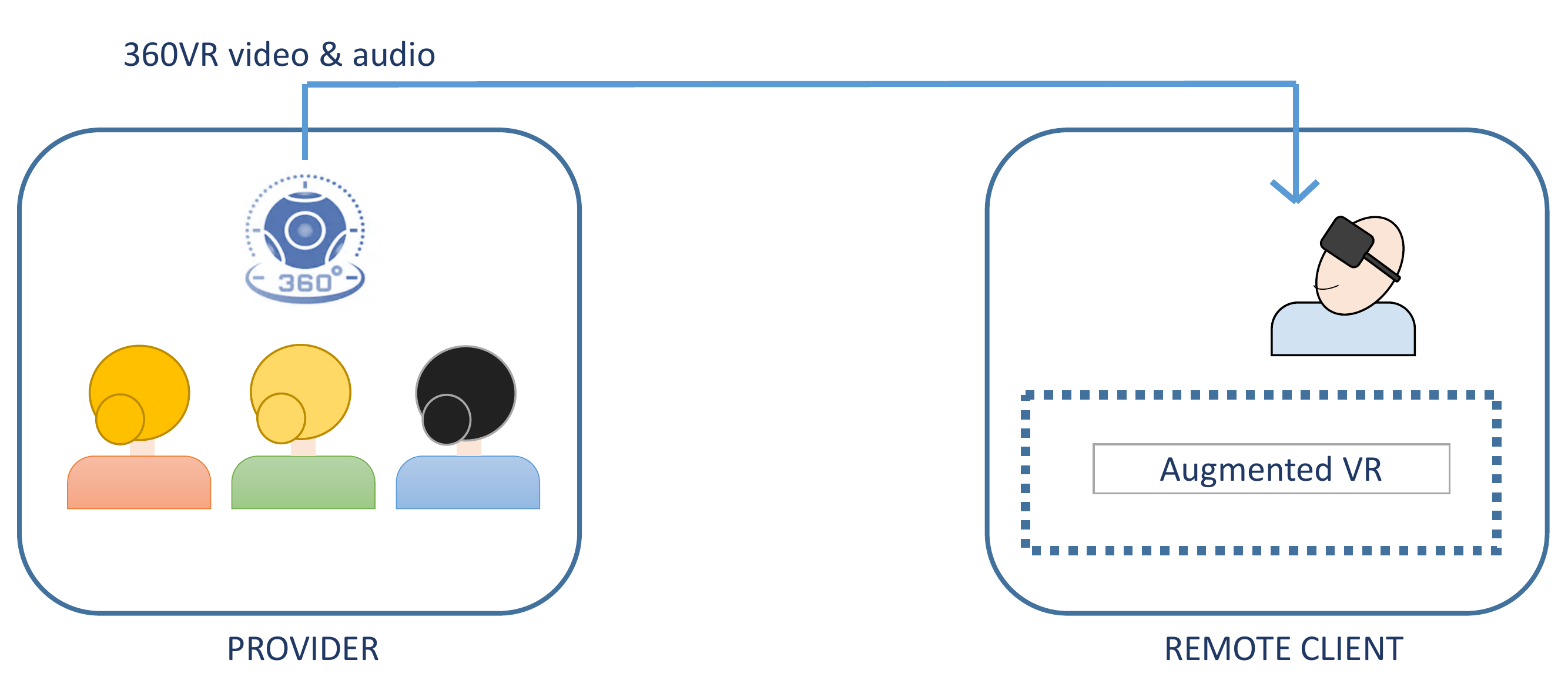}
  \caption{\moc{Simulated immersive communication environment of the experiment}}
\label{fig:Scenario}
\vspace{-3mm}
\end{figure} 

 The main reason is that 
 the use of real-time 360-degree video provides additional value as a communication platform that goes one step beyond traditional audio and video transmission~\cite{mocdcieee}. \moc{Typically, this type of platform is based on the architecture shown in Figure~\ref{fig:Scenario}, where 360-degree video and audio are transmitted from a particular location to a remote user, and are displayed by a Head-Mounted Display~(HMD). It allows the transmission of additional non-verbal signals such as facial expressions or body postures that are exchanged during a conversation, which greatly improve the effectiveness of face-to-face communications~\cite{grondin2019empathy}. On the other side, the remote user application displays some augmented information and provides interactivity using VR controllers~\cite{rhee2020augmented} or the user's hands~\cite{lee2018user, kachach2021immersive}. Besides, a return channel (not shown in the figure) normally exists, but its implementation varies significantly between different works, mainly depending on the use case. } 
 
 \moc{As a conclusion, VR provides more immersive environments and interactive experiences than today's communications technology to the user who attends the conversation with an HMD~\cite{thies2018facevr}. Thanks to these engaging environments, users can evoke psychological effects such as a sense of presence or other affective skills that we refer to as socioemotional features~\cite{salminen2019evoking}. }
 
In order to satisfy users' demands and expectations, it is essential to study and guarantee a high Quality of Experience~(QoE), which is affected by technical parameters but also by socioemotional features. \moc{In reference to the technical features, omnidirectional content is much more demanding than traditional content~\cite{itutg1035}. Specifically, a higher resolution than 2D videos is required to provide similar video quality due to the fact that the pixels are distributed in a 360-degree sphere. Also, higher framerates, ideally equal to the refresh rate of the HMDs used for the visualization, are required to avoid annoyance for participants. These requirements  dramatically increase bandwidth.  
In this sense,} several works in the literature have explored techniques to save bandwidth while offering acceptable video quality. Most approaches are based on non-uniform schemes, encoding and transmitting with higher quality only the field of view that the user is visualizing~\cite{ozcinar2019visual}. 
Other works are based on the fact that users tend to look at certain parts of the scene that are more attractive. In these cases, saliency or attention maps are computed to efficiently distribute the bitrate~\cite{david2018dataset}. 

\moc{The evaluation of the video quality achieved in these solutions or, in general, in the transmission of 360-degree video, is typically performed using subjective assessment tests based on methodologies highly proven with traditional contents~\cite{bt2012methodology, itutp910,p913}, and recently adapted to immersive video~\cite{9474501}}. Commonly, the stimuli used with these methodologies are short-duration videos without narrative, which are randomly displayed in different qualities and consumer devices to be rated. The problem is that this kind of methodologies has not been designed taking into account the requirements of socioemotional aspects. The analysis of empathy, spatial and social presence, attention or other affective skills that are experienced in a VR scenario requires long-duration videos with a narrative and genre adapted to the purposes of the experiment~\cite{schutte2017facilitating,att,tussyadiah2018virtual,fonseca2016comparison}. 
\moc{The whole purpose of using VR for teleconferencing is benefiting from its higher immersion and sense of presence; if the quality evaluation test does not provide such features, then its results might be not valid for the desired use case.}

Considering the current situation with the COVID-19 pandemic and the relevance of teleconferencing scenarios, VR technology can foster a change in communications. However, it is necessary to further investigate and provide standardized methodologies to consider all aspects that influence the QoE for the final boost of this technology~\cite{techmeetspsycho,perkis2020qualinet}. \moc{ Once there is literature that analyzes different socioemotional and technical aspects, an important advance should be to evaluate them together in experiments closer to real scenarios and use cases, increasing ecological validity and reliability. In addition, experiments that consider aspects that have already been independently evaluated saves time and resources. So, in this paper, we not only present an experiment with a methodology designed to evaluate both technical and socioemotional aspects; we launch a renewed point of view: how the evaluation of technical aspects influences socioemotional aspects, and vice versa, accelerating immersive communications as a solution to the current situation.} 

\subsection{Contributions}
This paper contributes to the fields of affective computing and quality of experience, providing an experiment where video quality and socioemotional aspects are jointly addressed. Here, we present in detail our main contributions:
\begin{itemize}
    \item \textbf{Methodology}. We propose and validate a methodology to jointly assess video quality and presence, empathy, attitude, and attention in immersive communications. This methodology is a solution for experiments in a controlled environment but more realistic than those presented in the literature. We propose the use of Single-Stimulus Discrete Quality Evaluation~(SSDQE) method to measure the quality during the test session and the aggregate quality in a post-questionnaire using the Absolute Category Rating~(ACR) on the same five-grade scale. Spatial and social presence are evaluated with an aggregate score obtained from 5 items based on the literature and adapted to our experimental environment. \moc{The initial empathy is evaluated using the Interpersonal Reactivity Index~(IRI). The attitude is measured in pre-questionnaire and post-questionnaire designed using facet theory. Due to the reliability of the scale, we propose to use only the post-questionnaire.} The attention is addressed with three questions about the scene that have pass/fail answers.
    \item \textbf{Video quality assessment in immersive communications}. We propose and verify an assessment of video quality for immersive communications using long-duration videos specifically designed and acquired for the exploration of socioemotional contents.
    \item \textbf{Dataset}. We make publicly available a Student Experiences Around the World dataset~(SEAW-dataset) of 3 video sources~(stereoscopic raw format) designed and acquired specifically for the purposes of the experiment. \moc{During the recording we considered three genres and both actor and observer acquisition perspective in the same context, international experiences, working or studying in a foreign country.} Additionally, the questionnaires and the associated rates obtained from a diverse and balanced sample of 54 participants are provided.  
    
\end{itemize}

The rest of the paper is structured as follows. Firstly, an overview of related works is presented in Section~\ref{sec:RelatedWork}. \moc{Then, Section~\ref{sec:RQs} presents the research questions.} 
 Section~\ref{sec:exdesign} explains the main features of the experiment design: \moc{experimental conditions,} test material, methodology, scenario, test session, and observers. 
The experimental results are presented in Section~\ref{sec:results} and finally, Section~\ref{sec:conclusions} includes general conclusions.

\begin{table*}[t]
\centering
\caption{Summary of datasets of 360-degree videos and evaluations}
\label{tab:dataset}
\resizebox{\textwidth}{!}{
\begin{tabular}{ccccc}
\multicolumn{1}{c}{\textbf{Authors}}                        & \multicolumn{1}{c}{\textbf{Number of 360-degree videos}} & \multicolumn{1}{c}{\textbf{Participants}}    & \multicolumn{1}{c}{\textbf{Socioemotional aspect}}              & \multicolumn{1}{c}{\textbf{Additional data}} \\\hline
\multicolumn{1}{c}{Li~et al. \cite{Li2017}} & 73                              & \multicolumn{1}{c}{95 (56 female, 39 male )} & \multicolumn{1}{c}{Valence and arousal}                         & \multicolumn{1}{c}{Participants' rotational head movements} \\
Jun et al. \cite{jun2020stimulus}           & 80                                                  & 551 (247 female, 262
male, 2 other)                                      & Presence, arousal, simulator sickness  & Future use intention of the video. Tracking data of the VR headset                                                             \\
                                                            Corbillon et al. \cite{corbillon2017360}           & 70                                                  & 59 20\% female)                                      &   & Tracking data of the VR headset     \\
Lo et al. \cite{lo2017360}           & 10                                                  & 50 (52\% male)                                      &   & Tracking data of the VR headset. Image saliency map from videos     \\
David et al.~\cite{david2018dataset}           & 19                                                  & 57 (25 female)                                     &   & Tracking data of the head and eye movement. Statistics related to exploration behaviors     \\
Yang et al. \cite{yang20183d}           & 13                                                  & 30 (15 female, 15 male)                                      &   & Mean opinion scores
\end{tabular}
}
\end{table*}

\section{Related work}\label{sec:RelatedWork}
The studies conducted in the literature present limitations that influence the results and conclusions and should be considered for the design of the experiment. In this section, we present an overview of the works mainly related to the quality and socioemotional features assessment.

\subsection{Quality Evaluation}





One of the main features to take into account during the design of subjective experiments is the test content. Despite the increase in consumption and therefore the creation of 360-degree content, high technical requirements are necessary for this kind of experiments. 
For example, problems caused during video acquisition or post-processing (e.g., stitching errors) or audio artifacts can influence the quality evaluations and affect the understanding of the content narrative. 
 Another aspect to take into account when selecting 360-degree content is its characterization in terms of exploration properties~\cite{DeSimone2019}. Generally, contents can be classified as directed or exploratory. Directed videos can help the observer to guide attention in the scene. Although participants move freely around the scene in exploratory contents, most of them fully explore the whole scene (360-degree) in 20~seconds~\cite{jun2020stimulus}. Nevertheless, contents of long-duration 
 can improve the engagement and enhance the emotions of the participants~\cite{fonseca2016comparison, janowski2019evaluating}. In addition to the duration, the genre and context of the video influence the success of the research. Specific 
 genres of content should be considered based on the socioemotional features addressed in the experiment~\cite{jun2020stimulus}, e.g: horror stimulus to test fear~\cite{macquarrie2017cinematic}. Taking this into account, we examined some 360-degree datasets in the literature with different characteristics, summarized in Table~\ref{tab:dataset}. For example, Li et al.~\cite{Li2017} released a public database of 360-degree videos covering a wide range of arousal and valence. Also, Jun et al.~\cite{jun2020stimulus} published a dataset containing 80 videos that were used to investigate a set of socioemotional features with a sample of 551 participants. They provided video sources with the corresponding report ratings and head movements. In addition, there are several datasets created to analyze exploration behaviors of the users when watching the content, such as the ones from Corbillon et al.~\cite{corbillon2017360} and Lo et al.~\cite{lo2017360} providing also head-movement data, or the one from David et al.~\cite{david2018dataset} that includes both head and eye tracking data. Regarding quality evaluation, some annotated datasets have been published, mainly containing short-duration videos, such as the one from 
 Yang et al.~\cite{yang20183d}. 
 
 The use of short-duration videos is a common approach on audiovisual quality evaluation, which is supported by several international recommendations related to subjective quality assessment, such as ITU-T P.910 and P.913~\cite{itutp910,p913, 9474501}. In these recommendations, standard assessment methodologies 
 are proposed to subjectively evaluate the impact of typical video artifacts (e.g., coding degradations) and \moc{also the guidelines for data processing. 
 For each video artifact, the mean of the evaluations of the observers with the associated Confidence Intervals~(CI) are computed to analyze the distribution of the means and their cumulative frequency of appearance~\cite{itutp910}. In the case of video quality, means are called Mean Opinion Score~(MOS) and typically, are presented with the associated 95\%~CIs. As these methodologies have been highly tested in the literature with 2D video, the distribution obtained with the representation of the MOS with the associated CIs of video quality evaluations helps the researcher to validate her/his experiment. Generally, it is more difficult for observers to appreciate the differences between very high quality content. However, in the videos encoded with intermediate qualities, the observers are able to find differences, but when the video quality is very low and annoying artifacts appear, the ratings saturate.} These methodologies, which were originally designed for 2D video, have been used in 360-degree video experiments and, somehow, adapted to address the new perceptual factors involved in 
 VR~\cite{singla2020quality,mocieee}, such as simulator sickness or exploration behavior. However, there is a research line supporting that quality assessment should be done under the most realistic conditions when services and applications are addressed to end users~\cite{garcia2014quality, pinson2014new}.  
 
 
 Based on the constraints presented, mainly related with content, methodologies, and context, we decided the design and acquisition of the 360-degree contents taking into account the purposes and the final devices used in the experiment~\cite{macquarrie2017cinematic}. With this, we propose a quality evaluation on long-duration videos with a context that interests or affects the participant and with a genre selected according to the purpose of the research. Also, we choose an environment where the participant is isolated, facilitating the real world disassociation.




\subsection{Socioemotional aspects evaluation}

Due to the limitations of the traditional QoE assessments, a great effort has been made in the analysis of socioemotional features in VR. Riva et al.~\cite{riva2007affective} demonstrated the effectiveness of VR as an \textit{affective medium}, a medium able to elicit different emotions through the interaction with its contents. Furthermore, the study demonstrated that the perceived sense of presence, related to a sense of being in a place~\cite{slater1997framework}, influences the emotional state. Following this research line, many studies have already confirmed the ability of VR to create more immersive environments, improving the socioemotional features. For instance, Fonseca et al.~\cite{fonseca2016comparison} 
demonstrated the highest emotional involvement of the participants viewing two types of narrative 360-degree contents with an HMD. MacQuarrie et al.~\cite{macquarrie2017cinematic} 
obtained a significant improvement of enjoyment of users using the HMD. In addition, VR emphasizes the phenomenon called Fear of Missing Out~(FoMO)~\cite{fomo}, defined, in the context of VR, as the apprehension that others might be having rewarding experiences from which the user with the HMD is absent. Additionally, users can freely move around the virtual environment, selecting the most interesting area of the 360-degree scene to focus on. 
These factors (immersion, FoMo, user motion pattern) may influence the attention that users pay to the events and objects in the scene. 
Some works in the literature analyze methods for assessing attention in this kind of environment~\cite{tong2019action, att}. 

Several works go one step further analyzing the use of this technology for empathy purposes and even for behaviour change purposes. Empathy is defined as the ability to view the world from another person's perspective combined with an emotional reaction to that perspective, including feelings of concern for others~\cite{davis1983measuring}. These studies are based on the fact that involvement created by VR environments facilitates empathy for users and can be used for specific purposes~\cite{schutte2017facilitating}. Aitamurto et al.~\cite{aitamurto2018sense} evaluated the responsibility for resolving gender inequality visualizing a 360-degree content in which participants could choose to watch the narrative from the male or female character's perspective. 
 Likewise, Tussyadiah et al.~\cite{tussyadiah2018virtual} confirm the effectiveness of VR technology in shaping consumers' attitude and behavior for tourism purposes.

\begin{table}[t!]
\renewcommand{\arraystretch}{1.1}
\begin{center}
\caption{\moc{Technical specifications of the test material used in the pilot study~\cite{pilotrepository}}}
\vspace{-2mm}
    \label{tab:pilotvideos}
\begin{tabular}{cccc}
\multirow{1}{*}{\textbf{Source content}} & \textbf{Resolution} & \textbf{Framerate (fps)} \\ \hline
\textbf{Alento}   & 3840x1920     & 25         \\ 
\textbf{AngelFalls}   & 3840x2160     & 30          \\  
\textbf{Flamenco}   & 3840x2160     & 30         \\ 
\textbf{LionKing}   & 3840x2048     & 30          \\ 
\textbf{Lions}   & 3840x1920     & 30          \\ 
\textbf{SwissJet}   & 3840x1920     & 50          \\ 
\hline
\end{tabular}
\vspace{-5mm}
\end{center}
\end{table}

\moc{Most of the literature can be divided into two main areas. One area focuses on the analysis of specific socioemotional features or a small subset of these independently. 
Many of those works 
do not address technical features such as resolution, framerate, or encoding parameters of the video~\cite{aitamurto2018sense}. The other area focuses mainly on technical parameters and only some socioemotional aspects are evaluated~\cite{schmidt2021towards,vlahovic2021effect}. We propose a methodology, following the experience of the literature, to assess video quality and several socioemotional features in the same experiment, reporting technical features, questionnaires, and sample diversity.
}



\subsection{Pilot study}
To further examine the findings from the literature, we conducted a pilot study where the influence of the HMD, usability, and fatigue in 360-degree video quality assessments were examined~\cite{mocieee}. The equipment used in the experiment consisted of two of the most popular HMDs with different evaluation methods, Samsung Galaxy S8 with Samsung Gear VR which includes a touchpad on its right side, and Lenovo Mirage Solo with a handheld controller. Regarding the stimuli, \moc{Table~\ref{tab:pilotvideos} presents the technical specifications of the} six representative sources \moc{with audio selected for the experiment. As recommended in ITU-T P.910~\cite{itutp910}, they cover a wide range of characteristics in terms of spatial and temporal information. Additional information is provided in the repository~\cite{pilotrepository}.} 
Clips of 25~seconds from the sources were encoded with ITU-T H.265/High Efficiency Video Coding~(HEVC) using fixed Quantization Parameters~(QPs): 22, 27, 32, 37, and 42~\cite{hmtestcond}. Video quality was evaluated using the ACR-HR~(Absolute Category Rating with Hidden Reference) with a five-level rating scale, as recommended in ITU-T P.910~\cite{itutp910}. Presence was assessed with two of the highly tested questionnaires: the Temple Presence Inventory~(TPI)~\cite{lombard2009measuring} and the Presence Questionnaire~(PQ)~\cite{witmer1998measuring}. As a result of this work, we provided a repository that contains\footnote{\moc{https://www.gti.ssr.upm.es/data/360VR}}: 
\begin{itemize}
    \item Dataset of video sources with the associated objective metrics results (PSNR, WS-PSNR, CPP-PSNR, VMAF, SSIM, MSSSIM) and details (Spatial and Temporal Indicators~\cite{itutp910}, resolution, framerates, \moc{and brief descriptions}).
    \item Head tracking data and video quality rates obtained from 48 participants during free-viewing experiments with two HMDs: Samsung GearVR and Lenovo Mirage Solo.
    \item Presence questionnaire scores, specifically TPI (Lombard et al.) and PQ (Witmer \& Singer), obtained from 48 participants.
    \item Statistical analysis notebook.
\end{itemize}

\moc{Following the literature, we corroborate that it is difficult to evaluate socioemotional aspects in short-duration clips where there is neither narrative nor context. In addition, the fact of repeatedly visualizing the same clips in different qualities and with two devices, made the participants initially evaluate the aspects related to presence in a positive way but nevertheless, as the experiment session progressed, the sense of presence decreased notably, what we call as fatigue effect. Additionally, some participants after the session told the researcher responsible for the experiment that the presence was highly dependent on the content. As we had not collected this information in a structured way, we considered in the experiment that we present in this paper higher-level aspects such as acquisition perspective, camera location, and interactive elements that could influence socioemotional aspects. The fact that the fatigue and higher-level aspects may affect the evaluation of socioemotional features is a huge motivation for the methodology for video quality evaluation proposed in this experiment. Also, the results of the pilot study help us to select the handheld controller as an evaluation method to increase the comfort of the observers. 
}
Additionally, we present a comparison of the scores obtained during the video quality evaluation of the pilot study and the experiment presented in this paper.


\section{Research Questions}\label{sec:RQs}


Based on the previous analysis, we pose the following Research Questions~(RQs):

\begin{itemize}
    \item RQ1: Is it possible to evaluate video quality in videos of long-duration designed for the evaluation of socioemotional features?
    \item RQ2: Which technical aspects, such as the position of the camera, the type of conversation, the video quality or the acquisition perspective influence socioemotional features? 
    \item RQ3: Which interactive elements can be provided to the remote client to improve some socioemotional aspects such as presence or attention? 
\end{itemize}

To answer these RQs, we designed a subjective experiment where an immersive communication between a \textit{provider} and a \textit{remote client} was simulated, presented in Figure~\ref{fig:Scenario}.
At the provider side, a conversation among several people took place, and the remote client attended virtually wearing an HMD. In the subjective test, the observer took the role of the remote client and visualized pre-recorded 360-degree videos with
fluctuations of quality, simulating a VR streaming communication.

The contents used in the experiment showed simulated conversations around a common topic: international experiences, i.e. working or studying abroad. 
The main idea behind choosing this specific context was our ability to gather a balanced sample of people who have had international experiences 
and with people who have not. We acquired \moc{360-degree videos }
 with different acquisition perspectives (actor and observer) and 
genre (everyday conversation, educational, and discussion). 
For that, student volunteers were recruited for the recordings, both exchange and national students from the university, making the conversations more realistic and fluent. 
Conversations were in English, making the experiment accessible to different nationalities and mother tongues and increasing the diversity of the sample.

\begin{table}[t]
\renewcommand{\arraystretch}{1.1}
\begin{center}
\caption{\moc{Overview of the three experimental conditions with the associated interactive element and features assessed in the experiment}}
    \label{tab:conds}
\begin{tabular}{cccc}
\multirow{2}{*}{\textbf{Condition}} & \multicolumn{2}{c}{\textbf{Assessment}} & \textbf{Interactive element} \\ 
\cline{2-4} 
              & \textbf{Quality} & \textbf{Socioemotional} & \textbf{Hands}\\ \hline
\textbf{A}   & X     & X      &       \\ 
\textbf{B}   &       & X      &       \\ 
\textbf{C}   &       & X      & X     \\
\hline
\end{tabular}
\end{center}
\end{table}

\begin{figure}[t]
  \includegraphics[width=0.5\columnwidth]{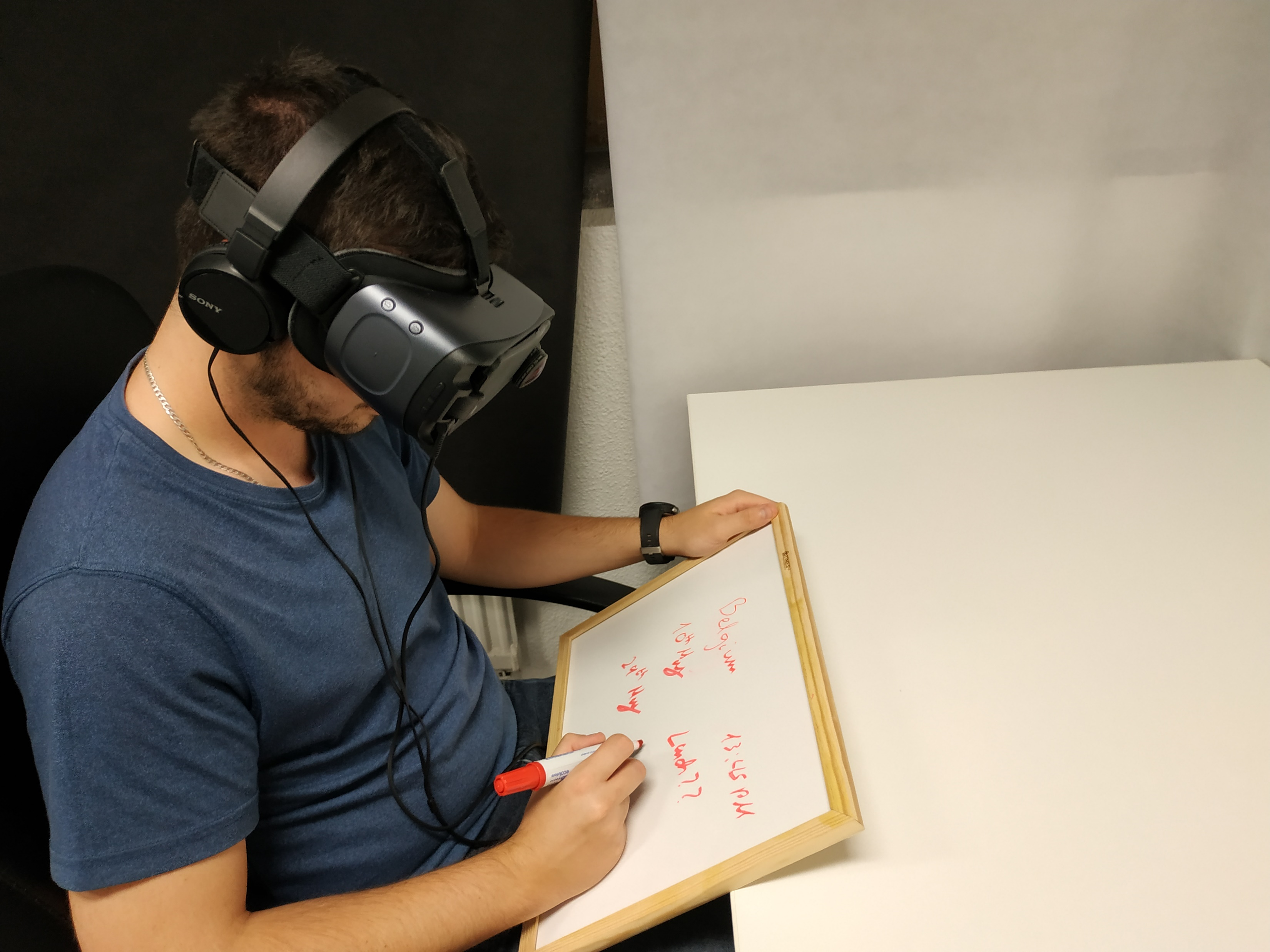}
  \centering
  \caption{Participant of condition C at the environment of the experiment}
  \label{fig:environment}
\end{figure}

\begin{figure*}[t]
    \centering
    \subfloat[\textit{Coffee shop}]
    {
        \includegraphics[width=0.33\textwidth]{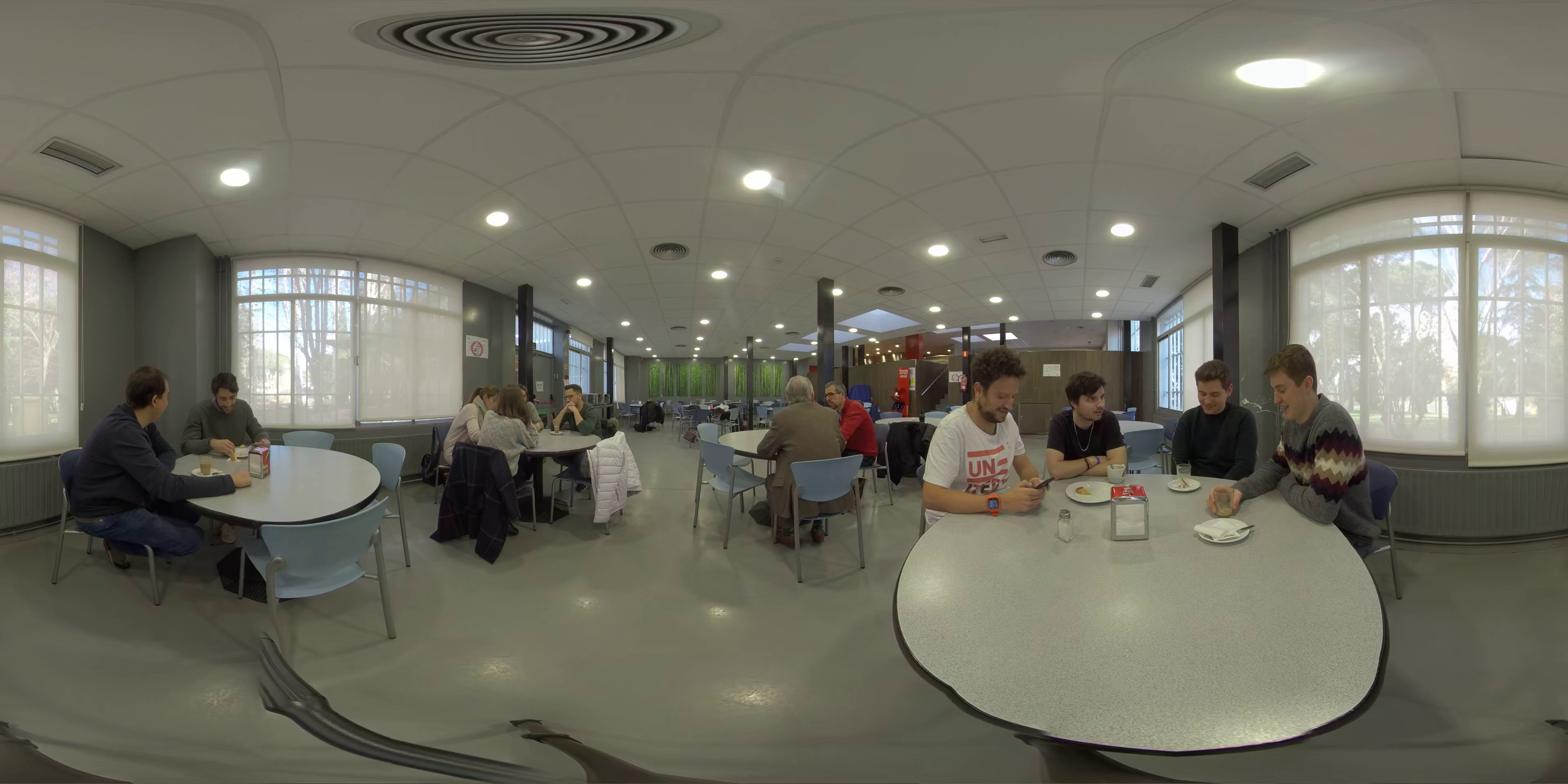}
    }
    \subfloat[\textit{International office}]
    {
        \includegraphics[width=0.33\textwidth]{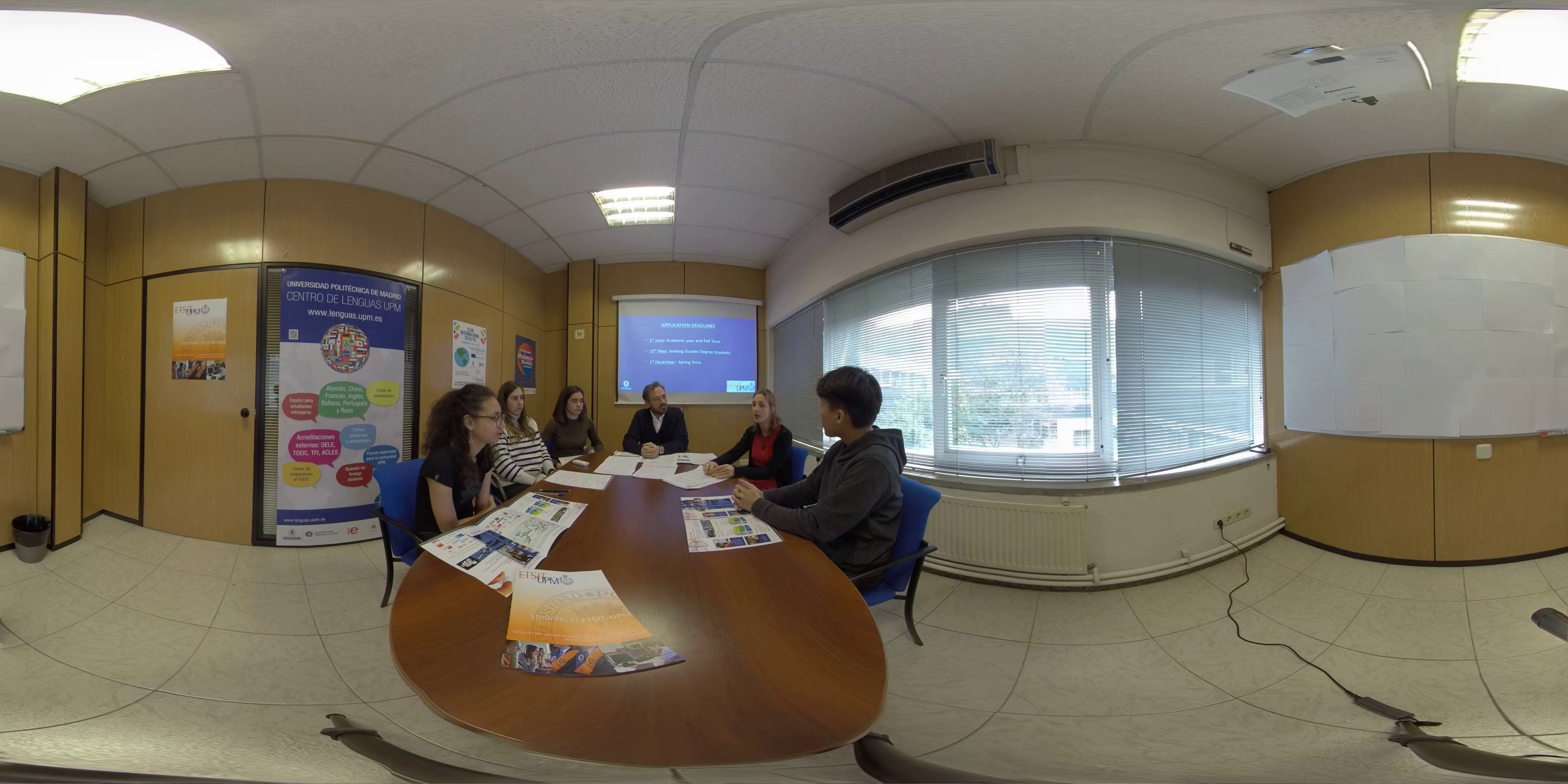}
    }
    \subfloat[\textit{Study in Spain}]
    {
        \includegraphics[width=0.33\textwidth]{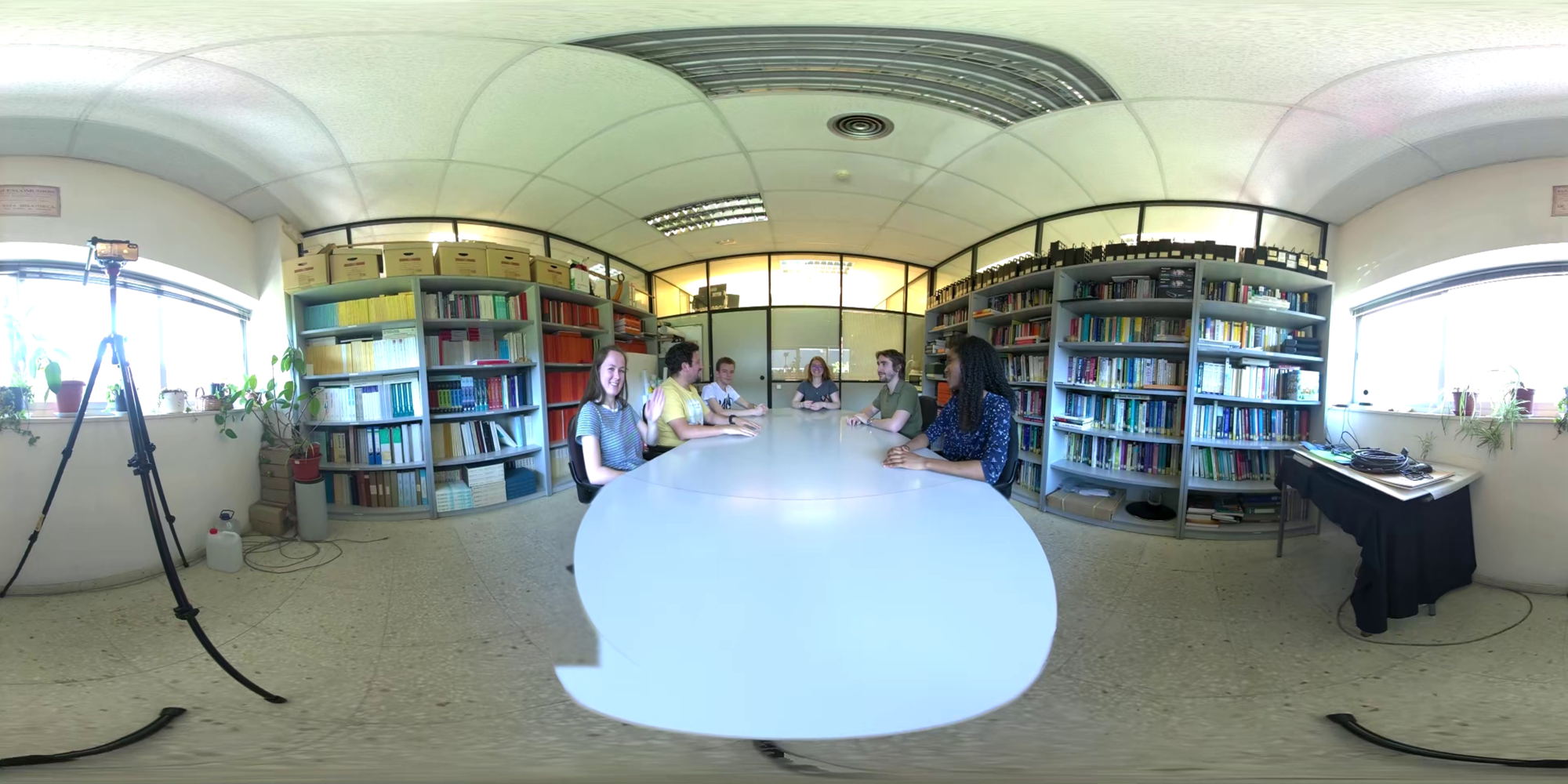}
    }
    \caption{Video sources screenshots}
    \label{fig:srcs}
\end{figure*}

\begin{table*}[t]
\renewcommand{\arraystretch}{1.2}
\caption{\moc{Characteristics of the 360-degree videos considered in the experiment}}\label{tab:material}
\centering
\begin{tabular}{cccc}
\textbf{Name}     & \textbf{Genre}        & \textbf{Perspective-taking} & \textbf{Description}                                \\\hline 
\textbf{Coffee shop}          & Everyday conversation & Observer                    & \textit{\begin{tabular}[c]{@{}c@{}}A coffee conversation between foreign and \\ local students about cultural differences\end{tabular}}           \\
\textbf{International office} & Educational           & Actor                       & \textit{\begin{tabular}[c]{@{}c@{}}A presentation given by a professor to\\  students about the foreign application process\end{tabular}}           \\
\textbf{Study in Spain}        & Discussion            & Actor                       & \textit{\begin{tabular}[c]{@{}c@{}}A conversation about the differences between \\ transport and rental prices in different countries\end{tabular}}
\end{tabular}
\end{table*}

\section{Experiment Design}\label{sec:exdesign}
The experiment was designed to jointly assess socioemotional features such as the sense of presence, empathy, attitude, and attention with video quality in a specific use case: a 360-degree communication. 

\subsection{Experimental Conditions}\label{sec:Conditions}

\moc{The experiment considered three test conditions, summarized in Table~\ref{tab:conds}, and each participant was assigned a condition. However, in all conditions, participants visualized the same video, 
with the same fluctuations of the quality. After each video, they were requested to rate its visual quality, as well as to evaluate the socioemotional features of interest: empathy and attitude, spatial and social presence, and attention.}

\moc{Participants assigned to condition A had the additional task of periodically rating the visual quality of the video during its playback, whenever its quality changed. This is a conventional design to evaluate the subjective quality of the video sequence under different intensities of impairment. However, this focused task might have impact on the evaluation of socioemotional features compared to the baseline scenario without the task (condition B).}

\moc{Finally, participants in condition C were provided with an additional interactivity element: the possibility to see their own hands and take handwritten notes about the conversation, as shown in Figure~\ref{fig:environment}. We hypothesize that this could enhance socioemotional features such as presence and attention with respect to the other conditions.}

\begin{table*}[t]
\caption{Structure of the test session questionnaires}
\label{tab:questionnaires}
\resizebox{\textwidth}{!}{%
\begin{tabular}{cccccccccc}
\multirow{2}{*}{\textbf{Condition}} & \multicolumn{3}{c}{\textbf{Pre-questionnaire (once)}}                                         & \textbf{During each content} & \multicolumn{5}{c}{\textbf{Post-questionnaire (for each content)}}               \\ \cline{2-10} \\
                            & \begin{tabular}{c}Personal \\ information\end{tabular} &  \begin{tabular}{c}Empathy \\ (IRI)\end{tabular} & \begin{tabular}{c}Attitude \\ \moc{(EM1-EM4)}\end{tabular} & \begin{tabular}{c}Quality \\ (SSDQE)\end{tabular} & \begin{tabular}{c}Quality \\ (ACR) \end{tabular} & \begin{tabular}{c}Attention \\ survey \end{tabular} & \begin{tabular}{c}Attitude \\ \moc{(EM5-EM8)} \end{tabular} & \begin{tabular}{c}Spatial Presence \\ \moc{(PP1-PP5)} \& \\ Social Presence \\ \moc{(SP1-SP5)}\end{tabular} & \begin{tabular}{c}Notes\end{tabular} \\ 
\textbf{A}                           & X & X & X & X                & X & X & X & X &                            \\ 
\textbf{B}                           & X & X & X &                  & X & X & X & X &                            \\ 
\textbf{C}                           & X & X & X &                  & X & X & X & X & X

\end{tabular}
}
\end{table*}

\subsection{Test Dataset}\label{Material}
The set of 
source videos, Student Experiences Around the World dataset~(SEAW-dataset) 
\moc{consists of} three stereoscopic contents in 4K resolution at 30~fps and a duration of approximately five minutes each were acquired and prepared specifically for the experiment. Figure~\ref{fig:srcs} shows a screenshot of the source videos and the original ones can be found in the supplementary material\footnote{https://www.gti.ssr.upm.es/data/seaw-dataset} 

As it can be observed in all sequences, student volunteers were sitting around a table far enough from the camera to avoid stitching problems affecting the user's QoE and video quality scores. In addition, the camera was placed at the position and average height of the head of a person sitting at the same table, facilitating the engaging experience~\cite{keskinen2019effect}. Table~\ref{tab:material} summarizes the genre, perspective-taking, and a brief description of the contents used in the experiment. In contents with the actor acquisition perspective, student volunteers during the recording looked at the camera, and even waved their hands to increase the immersion of the participant of the experiment visualizing the 360-degree content with the HMD.  

The \moc{contents} 
were encoded with HEVC switching to a different fixed QP each 25 seconds to create one \moc{Processed Video Sequence~(PVS)} per source content~\cite{bt2012methodology}. The QPs selected for the experiment were: 15, 22, 27, 32, 37, and 42~\cite{hmtestcond}. These QPs were randomized along the video encoding, following Rec. ITU-R BT.500-13~\cite{bt2012methodology}. Based on the assumption that each video source maintains the features in terms of color, texture, composition, and light, participants rated the quality of each one of the 25-second units along the whole sequence, avoiding the repetition of the same clip~\cite{janowski2019evaluating}. Due to the duration of the contents, each QP was rated at least two times in each PVS. Finally, the original audio quality was maintained through the experiment, improving the immersion and the QoE of the observers~\cite{tse2017there}.

\subsection{Methodology}\label{sec:methodology}
Here, we explain in detail the methodology considered in the experiment. Table~\ref{tab:questionnaires} 
summarizes the items evaluated in the three experimental conditions. 

\textbf{Personal information:} For each participant, we collected 
age, gender, vision (corrected or normal), nationality, experience living in a foreign country and which one, and English level. This was used to characterize our observers and guarantee diversity.

\textbf{Empathy}\label{empathy}. The initial empathy of each of the observers was evaluated using the Interpersonal Reactivity Index~(IRI)~\cite{davis1983measuring}. This questionnaire is a psychometrically invariant empathy measure based on 28 statements related to the Perspective-Taking scale~(PT), Fantasy Scale~(FS), Empathic Concern scale~(EC), and Personal Distress scale~(PD). For each statement, the observer was required to indicate how well it described her/him on a five-level scale~(where 1 = "Does not describe me well", to 5 = "Describes me very well").

\textbf{Attitude}\label{attitude}. A survey was designed to measure the attitude towards the context of the videos, international experiences. As there was no validated questionnaire to measure the attitude of the participants towards other cultures and foreigner experiences, we decided to apply the Facet theory~\cite{borg2005facet}. Facet theory consists of distinguishing the facets in which the designers of the experiment are interested. From the identified facets, the items of the questionnaire are defined and associated. In our case, we identified four characteristics that a person with a positive attitude towards foreigners and other cultures must have: interest, tolerance, respect, and social sensitivity. We established four statements related to the \textbf{interest}
, \textbf{respect}, 
\textbf{tolerance,} 
 and \textbf{social sensitivity} towards other cultures and traditions.
These four items were evaluated before starting the session and after the visualization of each of the three videos analyzed in the experiment. In this way, we could compare the empathy and attitude evolution throughout the session. To do this, four questions (EM1-EM4) were designed for the first evaluation at the beginning of the test, and another four questions (EM5-EM8) for the evaluation after each video. \moc{The idea behind this design was to compare the ratings before the visualization of the 360-degree content and after it. }
 Observers provided ratings on a seven-point Likert scale~(where 1 = "Strongly disagree", to 7 = "Strongly agree") based on most works in the literature~\cite{aitamurto2018sense, mocieee}. Specifically, it was measured with the following questions: \textit{ I just need to know the traditions of my country of origin~(EM1), I think that some traditions of other cultures should not be allowed in my country~(EM2), I feel comfortable with traditions different from mine~(EM3), and I am worried about the experiences of foreign people in my country~(EM4), I like participating in this kind of conversation~(EM5), I think that an intercultural society has a positive impact for the people~(EM6), I would feel comfortable sharing traditions of other cultures~(EM7), I would like to participate in a buddy program or in a project to know more about foreign experiences in my country~(EM8).}

\begin{table*}[t]
\renewcommand{\arraystretch}{1.5}
\centering
\caption{\moc{Attention survey: True/False statement, short answer, and multiple choice question for each of the 360 videos}}\label{tab:attq}
  \begin{tabular}{ >{\centering\arraybackslash}m{2cm} >{\centering\arraybackslash}m{4cm} >{\centering\arraybackslash}m{4.5cm} >{\centering\arraybackslash}m{5.5cm} }
  \textbf{Content} & \textbf{True/False statement} & \textbf{Short answer question} &  \textbf{Multiple-choice question} \\
  \hline
  \textbf{Coffee shop} & Students think that the second year is easier & How long ago has the university education system in Spain changed? & What city are the exchange students from? \\
  \textbf{International office} & All students can apply for an internship both in research groups and in companies & The deadline to apply for double degree students & The deadline to apply for an internship or exchange program for the whole year \\
  \textbf{Study in Spain} & Norwegians spend on average less money on public transport & The price of the public transport card in France & The rent per month in Norway \\
  \end{tabular}
  \label{tab:TestConditions} 
\end{table*}

\begin{figure}[t]
  \includegraphics[width=1\columnwidth]{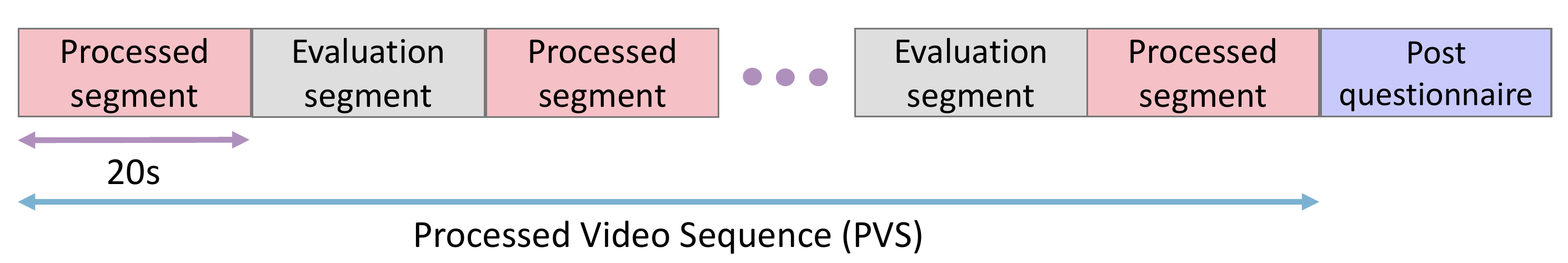}
  \caption{Structure of the test sequences used with SSDQE methodology}
\label{fig:SSDQE}
\end{figure}

\textbf{Quality}\label{Quality}.
A Single-Stimulus Discrete Quality Evaluation~(SSDQE) method~\cite{gutiarrez2012validation} was applied to measure the quality in observers assigned to condition A. SSDQE uses long-duration contents to evaluate quality guaranteeing the continuity of the narrative. For this, the content is divided into segments, trying to mimic realistic situations of video consumption. As represented in Fig.~\ref{fig:SSDQE}, impairments are inserted throughout the content used as stimuli (PVS) in alternate segments ("processed segments") and participants rate the perceived quality during the following ones ("evaluation segments"). \moc{Note that during the evaluation segment, video playback continues and is encoded with the same quality as the previous processed segment.} Specifically, they evaluated the quality on a five-grade quality scale~\cite{bt2012methodology}, where the categories: "Bad", "Poor", "Fair", "Good", and "Excellent" were displayed on the screen. Additionally, the aggregate quality was asked, following the literature, in the post-questionnaire 
using the Absolute Category Rating~(ACR) on the same five-grade scale~\cite{gunkel2018virtual, covaci2019360}.

\textbf{Attention}\label{Attention}.
Observer attention was assessed with three questions about the conversations taking place in the videos that had pass/fail answers~\cite{att,macquarrie2017cinematic}. For each content, we designed a multiple-choice question, a short answer question, and a True/False statement, presented in Table~\ref{tab:attq}. In this way, participants scored zero or one point for each correct answer, resulting in a maximum score of three points for the total attention score for each video.

\textbf{Presence}\label{Presence}.
Spatial and social presence experienced by the observers were evaluated with five questions obtained from the state of the art~\cite{riva2007affective, aitamurto2018sense}. \moc{The questions of social presence questionnaire were mainly related to factors that influence the involvement in the meeting, such as the feeling that people in the meeting are looking at us, talking to as or where is the group attention focused on~\cite{slater1997framework}. } Observers provided ratings on a seven-point Likert scale~(where 1 = "Strongly disagree", to 7 = "Strongly agree"). Specifically, spatial presence was measured with the following questions: \textit{I felt I was present in the places shown in the video~(PP1), I felt surrounded by the actions in the video~(PP2), I felt I was sitting by the table at the place of the video~(PP3), I felt I could have reached out and touched the items on the table of the video~(PP4), and I felt that all my senses were stimulated at the same time~(PP5).} Likewise, social presence was measured with the following ones: \textit{I felt that people were talking to me~(SP1), I felt that I was listening to the others in the video~(SP2), I felt I was present with the other people in the video~(SP3), I felt like the people in the video could see me~(SP4), and I felt I was actually interacting with other people~(SP5).}

\textbf{Notes}: Participants assigned to condition C assessed the usefulness of the notes taken during the test session. Specifically, the question \textit{Have your annotations helped you to solve the questions?} was used to consider whether the correct answers were correct from the annotations or from the memory of the participants.


\begin{figure}[t!]
  \includegraphics[width=1\columnwidth]{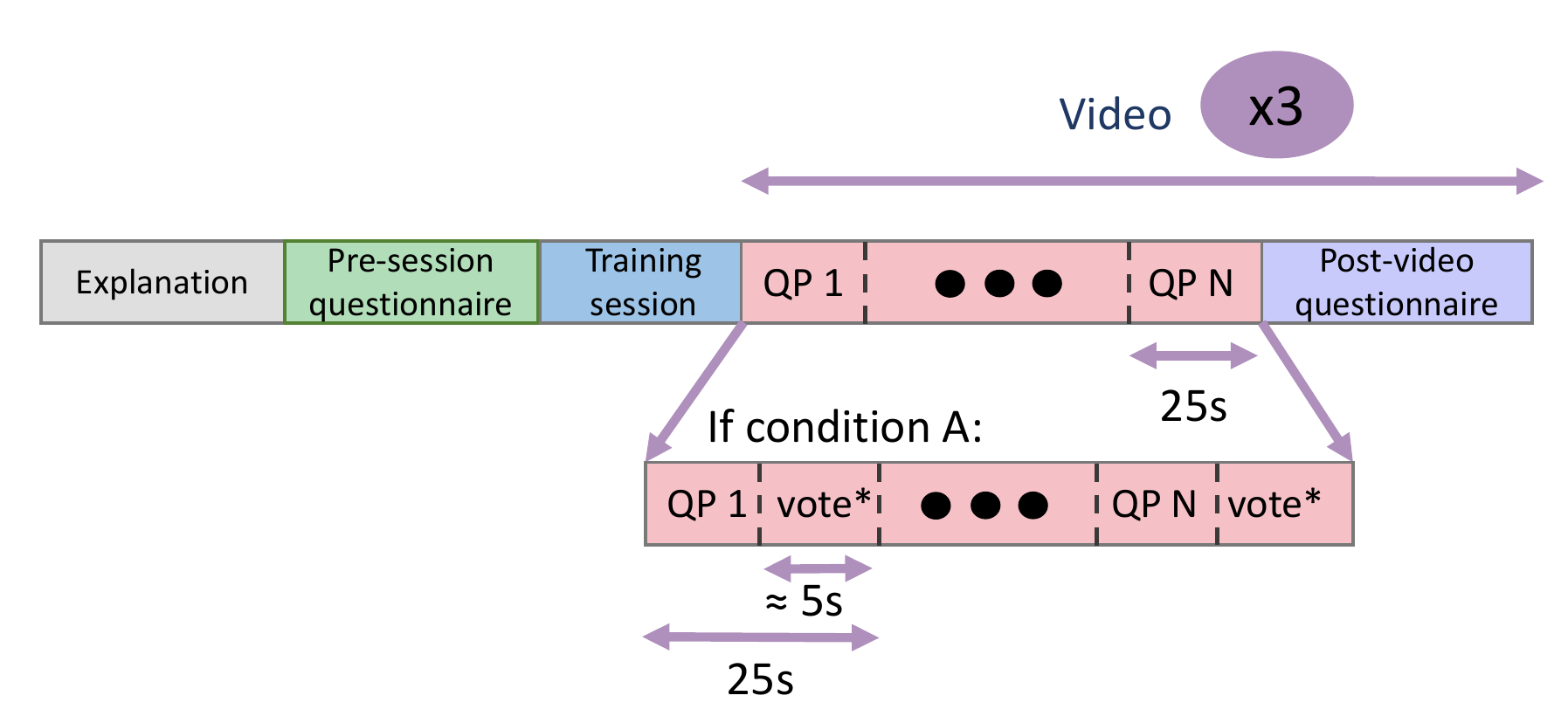}
  \caption{Test session structure}
  \label{fig:test_session}
\end{figure}

\begin{figure*}[t!]
    \centering
    \subfloat[\textit{Condition A}]
    {
        \includegraphics[width=0.33\textwidth]{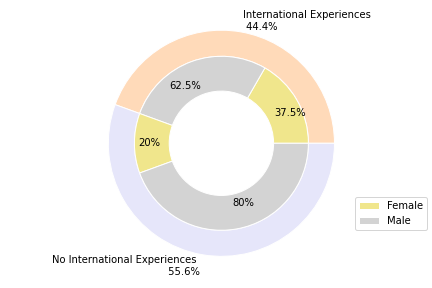}
    }
    \subfloat[\textit{Condition B}]
    {
        \includegraphics[width=0.33\textwidth]{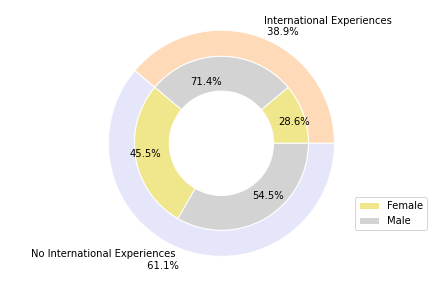}
    }
    \subfloat[\textit{Condition C}]
    {
        \includegraphics[width=0.33\textwidth]{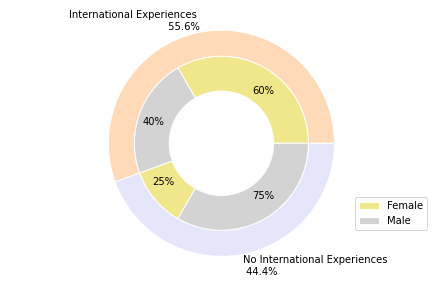}
    }
    \caption{Observers distribution in conditions A, B, and C taking into account the gender and international experience}
    \label{fig:pie_intgen}
\end{figure*}

\subsection{Environment and equipment}
All participants visualized the contents with a Samsung Galaxy S8 and the last model of Samsung Gear VR headset endowed with head tracking. The maximum resolution that viewers could perceive with this HMD (assuming a field of view of 85$^{\circ}$x100$^{\circ}$~and a smartphone native resolution of 1440x2960 pixels), is  about 680x822 pixels~\cite{munoz2020methodology}.
Monophonic audio was heard through headphones.

In all conditions, A, B, and C, the questionnaires  
were presented and answered using a web application. Observers who were assigned to condition A evaluated the quality of the video during the session. For this purpose, a VR application that allows users to visualize contents and answer customized questionnaires without having to take off their goggles was used~\cite{cortes2019unity3d}. They used a handheld controller as the evaluation method because it is more natural than the touchpad, avoiding any sign of discomfort~\cite{mocieee}. Observers assigned to condition C were able to see their own hands, as well as a small whiteboard to take some notes, using an Augmented Virtuality~(AV) approach, as shown in~\cite{perez2019immersive}. The local environment was captured by the smartphone camera and displayed in front of the 360-degree video. Background was removed from the camera image using chroma-keying based on red chrominance. 

Regarding the local environment, the observers were seated in a swivel chair in front of a table. This 
 chair allowed them to spin around without more limitations than the three degrees of freedom, imposed by the HMD. The table in front of them was a requirement imposed by the 
 videos, since, as presented in Figure~\ref{fig:srcs}, the three contents simulate a meeting around a table. In this way, observers could identify the table of the videos with the real one. Additionally, participants were located in totally isolated cubicles, facilitating the immersion in the content and avoiding any external distraction that increases the sense of FoMO.

\begin{figure}
\centering
\includegraphics[width=\columnwidth]{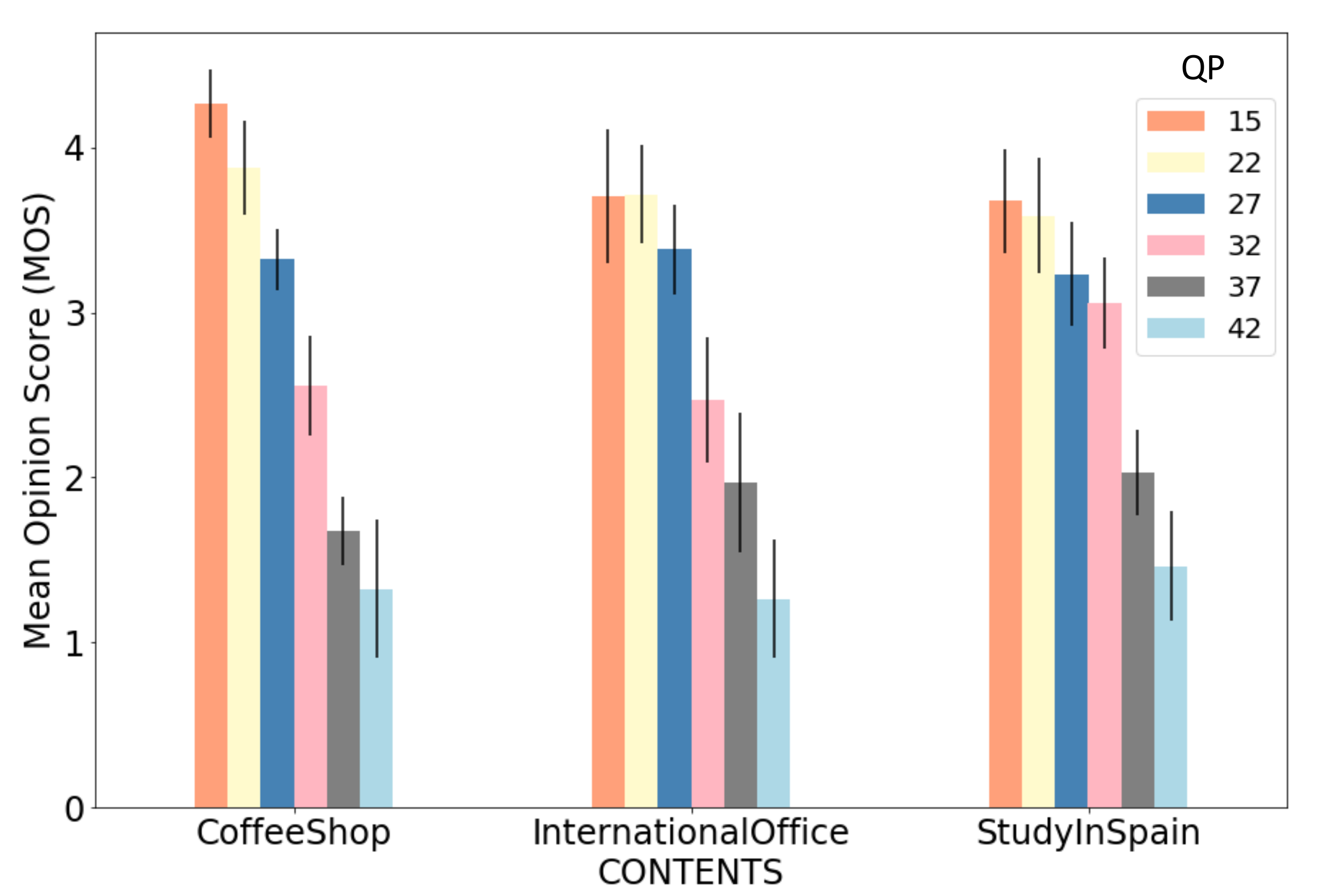}
\caption{\moc{The mean opinion scores (y-axis) on a five-level scale obtained from 17 participants assigned to condition A who evaluated the perceived video quality in the processed segment, encoded with specific QP, every 20~seconds while visualizing each of the three contents (x-axis), following the SSDQE methodology.} {\moc{Error bars represent 95\% CI.}}}
\label{fig:quality}
\end{figure}

\begin{figure}
\includegraphics[width=1.01\columnwidth]{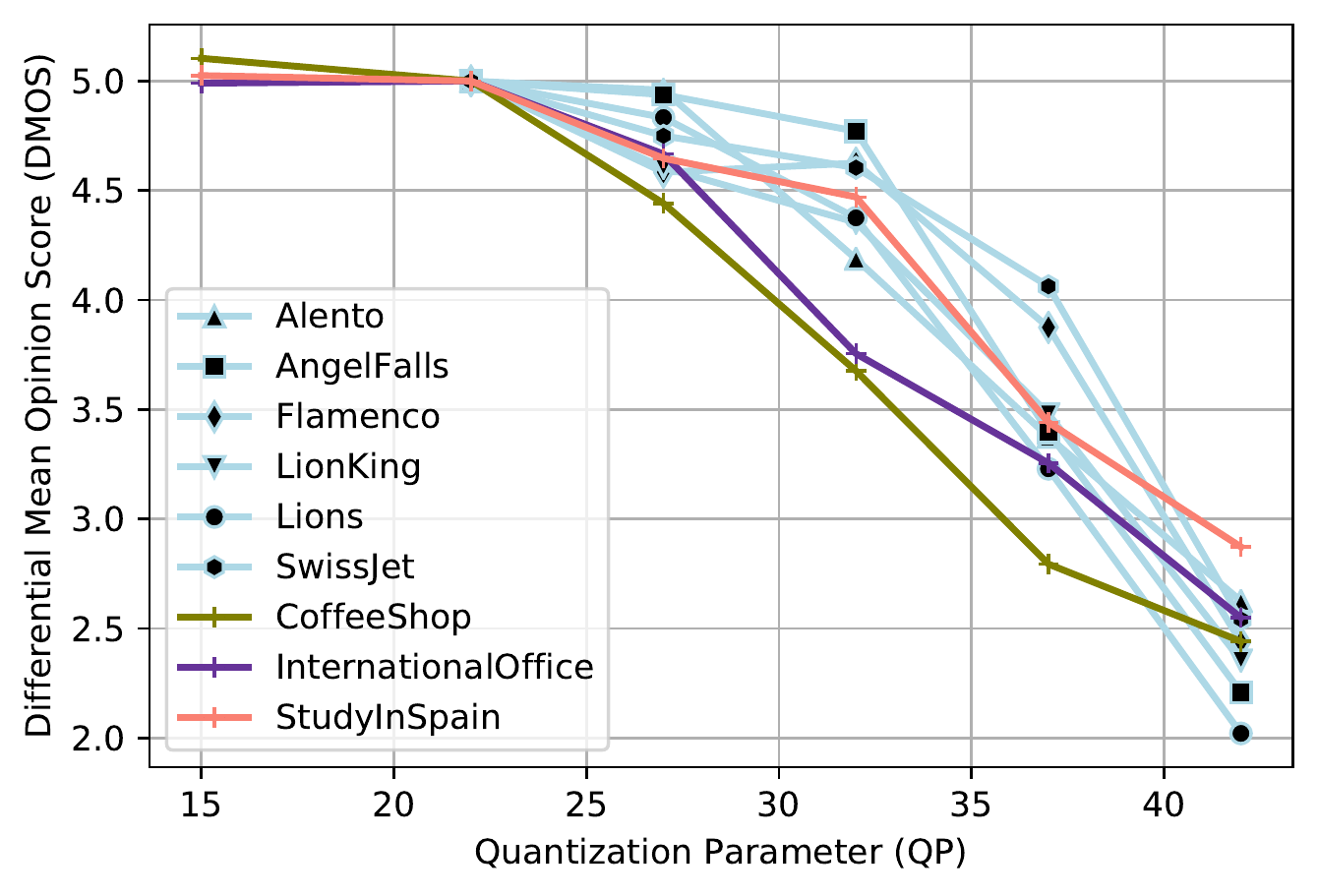}
\caption{\moc{Comparison of DMOS (y-axis) on a five-level scale obtained from 17 participants assgined to condition A, and from participants from the pilot study. Both participants evaluated clips of short duration encoded with fixed quantization parameters (x-axis). However, participants assigned to Condition A evaluated perceived quality following SSDQE methodology and participants from the pilot study evaluated it following ACR methodology.}}
\label{fig:quality_comp}
\end{figure}

\begin{table*}[t!]
\centering
\caption{\moc{Difference in aggregate quality and socioemotional features between the three conditions}}
\label{tab:conditions_dif}
\resizebox{\textwidth}{!}{
\begin{tabular}{ccccc}
\textbf{\begin{tabular}[c]{@{}c@{}}Questionnaire items \end{tabular}} & \textbf{Condition A}                                            & \textbf{Condition B}                                            & \textbf{Condition C}                                            & Significance                         \\ \hline 
\textbf{Aggregate quality 5-level scale}                                              & \begin{tabular}[c]{@{}c@{}}$M = 3.537$ \\ $(SD = .719)$\end{tabular}  & \begin{tabular}[c]{@{}c@{}}$M = 3.111$\\ $(SD = 1.022)$\end{tabular} & \begin{tabular}[c]{@{}c@{}}$M = 3.167$\\ $(SD = .885)$\end{tabular}  & $F_{2,153}=3.687, p<.05$, $\eta_{p}^{2}=.045$, $\gamma=.687$           \\ 
\textbf{Spatial Presence (7-level scale)}                                 & \begin{tabular}[c]{@{}c@{}}$M = 5.463$\\ $(SD = 1.019)$\end{tabular} & \begin{tabular}[c]{@{}c@{}}$M = 5.185$\\ $(SD = 1.318)$\end{tabular} & \begin{tabular}[c]{@{}c@{}}$M = 5.411$\\ $(SD = .942)$\end{tabular}  & $F_{2,153}=.900$, $p>.05$, $\eta_{p}^{2}=.011$, $\gamma=.203$      \\ 
\textbf{Social Presence (7-level scale)}                                 & \begin{tabular}[c]{@{}c@{}}$M = 5.059$\\ $(SD = 1.398)$\end{tabular} & \begin{tabular}[c]{@{}c@{}}$M = 5.133$\\ $(SD = 1.287)$\end{tabular} & \begin{tabular}[c]{@{}c@{}}$M = 5.144$\\ $(SD = 1.271)$\end{tabular} & $F_{2,153}=.005$, $p>.05$, $\eta_{p}^{2}=.000$, $\gamma=.05$       \\ 
\textbf{\moc{Attitude post-questionnaire} (7-scale level)}                               & \begin{tabular}[c]{@{}c@{}}$M = 5.875$\\ $(SD = .250)$\end{tabular}   & \begin{tabular}[c]{@{}c@{}}$M = 6.042$\\ $(SD = .108)$\end{tabular}  & \begin{tabular}[c]{@{}c@{}}$M = 6.236$\\ $(SD = .191)$\end{tabular} & $F_{2,153}=4.660$, $p<.05$, $\eta_{p}^{2}=.055$, $\gamma=.782$ \\ 
\textbf{Attention (3-level scale)}                                        & \begin{tabular}[c]{@{}c@{}}$M = 1.981$\\ $(SD = .765)$\end{tabular}  & \begin{tabular}[c]{@{}c@{}}$M = 1.833$\\ $(SD = .885)$\end{tabular}  & \begin{tabular}[c]{@{}c@{}}$M = 1.685$\\ $(SD = .748)$\end{tabular}  & $F_{2,153}=1.839$, $p>.05$, $\eta_{p}^{2}=.023$, $\gamma=.391$ \\
\end{tabular}%
}
\end{table*}

\begin{table*}[t!]
\centering
\caption{\moc{Difference in aggregate quality and socioemotional features between the three contents}}
\label{tab:contents_dif}
\resizebox{\textwidth}{!}{%
\begin{tabular}{ccccc}
\textbf{\begin{tabular}[c]{@{}c@{}}Questionnaire items\end{tabular}} & \textbf{\begin{tabular}[c]{@{}c@{}}Coffee shop\end{tabular}} & \textbf{\begin{tabular}[c]{@{}c@{}}International office\end{tabular}} & \textbf{\begin{tabular}[c]{@{}c@{}}Study in Spain\end{tabular}} & Significance                       \\ \hline
\textbf{Aggregate quality (5-level scale)}                                              & \begin{tabular}[c]{@{}c@{}}$M = 3.111$\\ $(SD = .904)$\end{tabular}  & \begin{tabular}[c]{@{}c@{}}$M = 3.222$\\ $(SD = .883)$\end{tabular}          & \begin{tabular}[c]{@{}c@{}}$M = 3.481$\\ $(SD = .885)$\end{tabular}    & $F_{2,153}=2.485, p>.05, \eta_{p}^{2}=.03$, $\gamma=.496$        \\ 
\textbf{Spatial Presence (7-level scale)}                                 & \begin{tabular}[c]{@{}c@{}}$M = 5.326$\\ $(SD = 1.173)$\end{tabular} & \begin{tabular}[c]{@{}c@{}}$M = 5.200$\\ $(SD = 1.137)$\end{tabular}         & \begin{tabular}[c]{@{}c@{}}$M = 5.533$\\ $(SD = .991)$\end{tabular}    & $F_{2,153}=1.394, p>.05, \eta_{p}^{2}=.017$, $\gamma=.297$   \\ 
\textbf{Social Presence (7-level scale)}                                  & \begin{tabular}[c]{@{}c@{}}$M = 4.748$\\ $(SD = 1.364)$\end{tabular} & \begin{tabular}[c]{@{}c@{}}$M = 4.752$\\ $(SD = 1.280)$\end{tabular}         & \begin{tabular}[c]{@{}c@{}}$M = 5.837$\\ $(SD = .964)$\end{tabular}    & $F_{2,153}=15.710, p<.01, \eta_{p}^{2}=.169$, $\gamma=1$  \\ 
\textbf{\moc{Attitude post-questionnaire} (7-scale level)}                               & \begin{tabular}[c]{@{}c@{}}$M = 6.111$\\ $(SD = .253)$\end{tabular}  & \begin{tabular}[c]{@{}c@{}}$M = 5.866$\\ $(SD = .234)$\end{tabular}        & \begin{tabular}[c]{@{}c@{}}$M = 6.176$\\ $(SD = .098)$\end{tabular}    & $F_{2,153}=3.271, p>.05, \eta_{p}^{2}=.038$, $\gamma=.605$   \\ 
\textbf{Attention (3-level scale)}                                        & \begin{tabular}[c]{@{}c@{}}$M = 2$\\ $(SD = .777)$\end{tabular}      & \begin{tabular}[c]{@{}c@{}}$M = 1.704$\\ $(SD = .743)$\end{tabular}          & \begin{tabular}[c]{@{}c@{}}$M = 1.796$\\$ (SD = .877)$\end{tabular}    & $F_{2,153}=1.925, p>.05$, $\eta_{p}^{2}=.024$, $\gamma=.407$  \\ 
\end{tabular}%
}
\end{table*}

\begin{table}[t]
\caption{\moc{Cronbach's $\alpha$ obtained for the questionnaires used in the experiment about spatial presence, social presence, and attitude}}
\centering
\begin{tabular}{cc}\label{tab:validation}
\textbf{Questionnaire}                 & \textbf{Cronbach’s }$\alpha$   \\ \hline
\textbf{Spatial Presence}              & 0.857           \\ 
\textbf{Social Presence}               & 0.865           \\ 
\textbf{Attitude (pre-questionnaire)}  & -0.094         \\ 
\textbf{Attitude (post-questionnaire)} & 0.710            \\ 
\end{tabular}
\end{table}

\subsection{Test session}
The test session structure is presented in Figure~\ref{fig:test_session}. At the beginning, participants received a brief explanation of the experiment. Also, they were informed and signed a consent form that allowed us to process the information in accordance with the General Data Protection Regulation~(GDPR) of the European Union. 
The experiment started with the pre-questionnaires: a personal information survey, the 
\moc{empathy questionnaire (IRI)}, and the initial attitude survey. The training session consisted of a visualization of a discussion around the table with a duration of two minutes approximately. The PVS used for the training sessions was encoded with the best and worst qualities offered in the experiment (QP values of 15 and 42) every 25~seconds.
 Observers assigned to condition A tested the evaluation method with the handheld controller. After the training session, the assessment session started. All participants visualized the same three PVS in a randomized order following Recommendation ITU-R BT.500-13~\cite{bt2012methodology}. 
For observers assigned to condition A, every 20~seconds the SSDQE question appeared without a time limit. 
After each video, all participants, regardless of the assigned condition, answered a post-questionnaire with questions about the quality and the socioemotional features. Here, participants assigned to condition C also answered the notes question. 

\subsection{Observers}
A total of 54 observers (20 females, 34 males) took part in this experiment. There were participants in the age range between 17 and 26 years, with a Mean age~(M) of 22.18 and a Standard Deviation~(SD) of 1.95. All observers were checked for normal or corrected-to-normal vision. All participants were required at least an intermediate level of English to understand the conversations of the videos. They received a small financial reward for participating. In this way, we obtained a sample of participants with international experiences or nationalities from 15 countries in Europe, America, and Asia. The representation of user diversity was an additional value of the experiment, since it increased its reliability~\cite{himmelsbach2019we,peck2020mind}. Furthermore, as it can be observed in Figure~\ref{fig:pie_intgen}, participants with international experiences and taking into account gender were distributed almost uniformly under conditions A, B, and C, guaranteeing a balanced sample. 

\section{Experimental Results}\label{sec:results}
For each one of the RQs, one or more hypotheses have been laid out and investigated, to look for relevant conclusions. Besides, the methodology to analyze the results was performed according to the nature of the data. The quality evaluation in condition A was examined with the \moc{MOS and the associated 95\% CIs} obtained from the scores, presented in Figure~\ref{fig:quality}. In regard to the quality and socioemotional features, the Pearson \& D'Agostino normality test was computed to validate the normal distribution of the collected data. For cases where the distribution was normal, the 2-way Analysis of Variance~(ANOVA) was performed to examine the differences among the evaluated videos and conditions. \moc{For social and spatial presence, due to the condition of non-normality, the following transformation was implemented: $\arcsin(\sqrt{P/7})$, where $P$ is the presence rating and it is divided by seven because social and spatial presence were evaluated in a seven-level scale. Once the data was transformed, it was analyzed under the normality condition.} 
 Post-hoc analyses using Bonferroni correction for multiple comparisons were applied to examine the differences among the evaluated videos and conditions. The considered level of significance was 0.05. Table~\ref{tab:conditions_dif} and Table~\ref{tab:contents_dif} present a summary of the scores of the items evaluated in the experiment and the significance \moc{($F$, $p$, partial eta-squared $\eta_{p}^{2}$, and observed power values $\gamma$)} between conditions and contents, respectively. 

\moc{
To investigate the factor structure of the questionnaires of presence and attitude, scale reliability has been addressed using Cronbach’s $\alpha$, presented in Table~\ref{tab:validation}. Considering a reliability scale with $\alpha>0.7$, the attitude responses from the pre-questionnaire have not been considered for the analysis. 
}

\subsection{Video quality assessment}
\label{subsec:vqa}

Regarding RQ1, we investigated the first hypothesis~(\textbf{H1)}: video quality evaluation can be adapted to long-duration videos designed for socioemotional features assessment purposes. 
 In this sense, we were interested in analyzing the effect of evaluating the quality of the video during the visualization of continuous sequences in which the scene features remain similar. Figure~\ref{fig:quality} was obtained from the scores of the 17 participants assigned to condition A. \moc{Note that the ratings of one of the observers were not collected correctly, so we remove an observer from condition A. As the evaluation of video quality on a 5-level score can be modeled by a Gaussian random process~\cite{janowski2015accuracy}, we use parametric analysis for the evaluation of the scores, following the common practices for video quality data evaluation~\cite{narwaria2018data}. We have performed a ANOVA to assess the dependency of the scores on each source video and each QP value. Results show that the QP is significant ($F_{5,596} = 186.4$, $p < .001$, $\eta^2 = .598$), while the source content is not ($F_{2,596} = 0.85$, $p > .05$, $\eta^2 = .018$). Bonferroni-corrected pairwise t-tests show that all pairs of QPs are significantly different between them, except the two higher qualities QP values of 15 and 22, which are not.} Note that due to the different duration of the videos and the randomization of the QPs, each QP was not evaluated the same number of times.

These quality scores were compared with the MOS values obtained from the pilot study presented previously~\cite{mocieee}, which was executed using a conventional ACR methodology with randomized 10-second video sequences. As the source contents were different in both experiments, we computed the Differential Mean Opinion Scores~(DMOS), according to ITU-T Recommendation P.910~\cite{itutp910}. We used QP 22 as the hidden reference, as it was the highest quality available in the pilot study, and it was shown not to be significantly different from QP 15 in our new experiment. Figure~\ref{fig:quality_comp} shows that both methodologies offer comparable results: good distribution of the ratings and a consistent decrease of the perceived quality when augmenting the QP, as expected in this type of tests~\cite{vmaf}.


These results show that subjects are able to effectively assess the video quality of individual QPs, and the content does not distract them from the task. This is in line with the results already reported in the literature for conventional 2D video and similar evaluation methodologies~\cite{gutiarrez2012validation,cranley2006user,ghadiyaram2017subjective}. Furthermore, having the subjects engaged in the content increases the ecological validity of the quality evaluation compared to traditional methods~\cite{pinson2014new, kortum2010effect}.

The aggregate quality scores rated at the end of each video in a five-level scale were analyzed statistically to find differences between videos and conditions. Due to the normality condition, 
2-way ANOVA was applied. Table~\ref{tab:contents_dif} shows that there is no significant difference between videos.
MOS lay somewhere in the middle between the lowest and highest scores obtained for individual QPs, which is also expected~\cite{tavakoli2016perceptual}. It is known that several factors, such as the amplitude, frequency, and time location of the quality switches have an effect on the formation of the overall quality opinion~\cite{garcia2014quality}, but addressing them is outside the scope of our experiment.

However, there is a significant difference among conditions, as seen in Table~\ref{tab:conditions_dif}. Student’s t-test with Bonferroni correction shows that this difference is significant between conditions A and B ($p=.0307$). Participants assigned to condition A 
scored the aggregate quality higher than participants assigned to condition B 
and C. 
It means that participants that are focused on the quality evaluation throughout the sequence, change their perspective about the perceived global quality. 

To the authors' knowledge, this result is new in the literature. Some authors have used similar methods to evaluate the video quality continuously during the content playback, and then a single \textit{endpoint quality} score at the end to assess the overall quality of the sequence~\cite{ghadiyaram2017subjective,bampis2017study}. \moc{However, none of them has also had the same sequences evaluated just at the end, as it is proposed, for instance, by ITU-T~\cite{raake2017bitstream}.} Our results show that the evaluation of quality during the sequence has a significant influence on the endpoint quality.

\subsection{Spatial and social presence}
\label{subsec:presence}
In reference to RQ2, we investigated the second hypothesis~(\textbf{H2)}: acquisition perspective, type of the conversation, and experimental condition have influence on: spatial and social presence.
 As said before, the items of the sense of spatial presence and social presence were measured on a seven-point Likert scale independently. As presented in Table~\ref{tab:conditions_dif} and Table~\ref{tab:contents_dif}, the analysis was twofold.
\moc{Once the non-normality condition of the social and spatial presence ratings was corrected, ANOVA was applied to analyze differences between experimental conditions}. The aggregate measure of the five spatial and social presence items respectively show that there is not a significant difference.  
 Nevertheless, a notable result is that the perceived social and spatial presence were very high in all conditions. 
In this sense, we want to point out that during the design of the experiment we presumed significant differences for condition C. We consider that the absence of differences is due to the fact that there were no specific tasks that required hands-on interaction with the VR environment.


\moc{Likewise, ANOVA was applied to examine the differences between videos.} The aggregate measure of the five items of spatial presence shows that there is no significant difference, but the aggregate measure of the five items of social presence does show a difference. 
\moc{Student's t-test with Bonferroni correction shows that there is a significant difference between "Study in Spain" and "Coffee shop" contents (${Z}=-4.887$, $p<.01$) and "Study in Spain" and "International office" content (${Z}=-5.023$, $p<.01$).}

Table~\ref{tab:contents_dif} shows that "Study in Spain" scored higher in 
social presence. To better explore the difference between contents, Wilcoxon Signed-Rank test with Bonferroni correction were applied to the items of the social presence questionnaire (SP1-SP5)~\cite{aitamurto2018sense}. The analysis shows significant differences between "Study in Spain" and the other 
videos, "Coffee Shop" and "International office", in questions related to the perception that people in the conversation speak, look at, and interact with the participant: SP1 (${Z}=47.5$, $p<.01$ and ${Z}=108$, $p<.01$), SP3 (${Z}=113.5$, $p=.0007$ and ${Z}=136$, $p=.015$), SP4 (${Z}=115.5$, $p<.01$ and ${Z}=108.5$, $p=.0001$), and SP5 (${Z}=107$, $p=.0005$ and ${Z}=86$, $p<.01$). The reason is that 
in "Study in Spain" content the actors appeal to the camera more frequently, emphasizing the non-verbal side of the conversation.



\subsection{Empathy and attitude towards international experiences}
\label{subsec:empathy}
Following with RQ2, we investigated the third hypothesis~(\textbf{H3)}: acquisition perspective, type of the conversation, and experimental condition have influence on: empathy and attitude.

\begin{table}[t]
\caption{\moc{The mean and standard deviations on a seven-level scale of the items of the attitude survey: interest, respect, tolerance, and social sensitivity in the three experimental conditions}}
\label{tab:attitude}
\resizebox{\columnwidth}{!}{
\begin{tabular}{ccccc}
\multirow{2}{*}{\textbf{Condition}} & \multicolumn{4}{c}{\textbf{Post-questionnaire}}                                                                                                                                                                                                                    \\ \cline{2-5} 
                                    & \textbf{Interest}                                               & \textbf{Respect}                                               & \textbf{Tolerance}                                             & \multicolumn{1}{l}{\textbf{Social Sensitivity}}                \\ \hline
\textbf{A}                          & \begin{tabular}[c]{@{}c@{}}$M = 5.167$\\ $(S = 1.411)$\end{tabular} & \begin{tabular}[c]{@{}c@{}}$M = 6.685$\\ $(S = .571)$\end{tabular} & \begin{tabular}[c]{@{}c@{}}$M = 6.407$\\ $(S = .806)$\end{tabular} & \begin{tabular}[c]{@{}c@{}}$M = 5.241$\\ $(S = 1.17)$\end{tabular}  \\ 
\textbf{B}                          & \begin{tabular}[c]{@{}c@{}}$M = 5.389$\\ $(S = 1.177)$\end{tabular} & \begin{tabular}[c]{@{}c@{}}$M = 6.574$\\ $(S = .71)$\end{tabular}  & \begin{tabular}[c]{@{}c@{}}$M = 6.333$\\ $(S = .861)$\end{tabular} & \begin{tabular}[c]{@{}c@{}}$M = 5.870$\\ $(S = 1.171)$\end{tabular} \\ 
\textbf{C}                          & \begin{tabular}[c]{@{}c@{}}$M = 5.537$\\ $(S = 1.343$\end{tabular}  & \begin{tabular}[c]{@{}c@{}}$M = 6.704$\\ $(S = .565)$\end{tabular} & \begin{tabular}[c]{@{}c@{}}$M = 6.407$\\ $(S = .913)$\end{tabular} & \begin{tabular}[c]{@{}c@{}}$M = 6.296$\\ $(S = .936)$\end{tabular}  \\ 
\end{tabular}
}
\end{table}

Firstly, IRI ratings were examined to obtain an adequate measure of the initial empathy of the participants, avoiding any deviation that may affect the subsequent analysis of the \moc{ attitude. }
 Given the condition of normality, ANOVA test was conducted to examine the IRI scores depending on gender and international experiences. It shows that there are not 
significant differences~($F_{1,50}=.76$, $p>.05$ and $F_{1,50}=3.838$, $p>.05$, respectively). Based on the literature~\cite{IRIEXAMPLE, gilet2013assessing}, we expected significantly higher scores for females than for males. In our case, there are not significant differences but on average females scored higher empathy than males both for participants with international experiences ($M=3.373$; $SD=.228$ and $M=3.316$; $SD=.237$) and for participants without international experiences ($M=3.246$; $SD=.302$ and $M=3.182$; $SD=.199$).   


Secondly, the attitude was evaluated 
\moc{with the questionnaire asked after the visualization of each of the three PVS.} Table~\ref{tab:attitude} summarizes the obtained results for each facet in the 
post-questionnaires~("Post"). \moc{Note that the data presented in the table is calculated in the original seven-level scale.}  
The attitude was measured with the aggregation of four items of the designed survey: interest, respect, tolerance, and social sensitivity. 
Table~\ref{tab:contents_dif} shows that there is not a statistically significant difference between contents but 
Table~\ref{tab:conditions_dif} presents a significant influence of the condition in which the content was visualized. 
Participants assigned to condition C achieved the highest attitude \moc{index}, followed by participants from condition B and A. After finding that the condition greatly influences on the attitude, Student's t-test with Bonferroni correction was applied to find differences between conditions. They show that the significant difference is only between A and \moc{C conditions (${Z}=-3.146$, $p=.002$).} 
It makes sense because participants assigned to condition A had the video quality assessment task, distracting them from the conversations taking place in the video. From this analysis, another main result is that there is an important positive impact in the three videos, and as presented, in the three conditions.

\subsection{Interactive element}

Finally, to answer RQ3 we investigate the fourth hypothesis~(\textbf{H4}): Observers who can take notes get higher total attention scores. 
Participants scored one point for each correct answer, resulting on a scale from 0 to 3. Due to condition of normality, ANOVA test was applied to find differences between conditions. As presented in Table~\ref{tab:conditions_dif}, the scores show that there is no a significant difference between participants assigned to condition A, B, and C. 
Among the 18 participants assigned to condition C, 10 of them reported that the notes they had taken helped them to answer the questions. However, their scores are not significantly different from the ones obtained by the other 8 participants.

\subsection{A method to simultaneously assess video quality and socioemotional features}

Our experiment shows that the methodology used for condition A is suitable for the simultaneous evaluation of video quality and socioemotional features. As shown in section~\ref{subsec:vqa}, SSDQE is valid to evaluate individual quality variations. Additionally, SSDQE does not affect the evaluation of presence or attention, which has two implications: on the one hand, it confirms that socioemotional features can be assessed despite having the extra task of continuous video quality evaluation; on the other, it shows that SSDQE does not reduce the observer immersion, making it a real content-immersive method.

There are, however, at least three caveats. First, using SSDQE does affect the evaluation of the overall quality of the sequence. Results obtained using this method will not be exactly the same as assessing the quality just with an endpoint evaluation. Second, as described in Section~\ref{subsec:empathy}, using SSDQE during the video has some impact 
on the attitude of the observers. This means that, although the simultaneously evaluation of quality and socioemotional features is possible, it is not completely neutral, and some interaction between evaluation tasks may exist. Finally, it is worth noting that the experiment has been done with a specific type of content and visualization (360-degree videos simulating conversations on international experiences). Other types of videos or visualization setups might have different behavior. Further research is needed to address these items.



    
\section{Conclusions}\label{sec:conclusions}

We have proposed a methodology where video quality, presence, empathy, attitude, and attention are jointly assessed in VR communications. We have simulated that user attend meetings remotely with the HMD and all meetings are focused on the international experiences context. Additionally, we have evaluated three conditions for the attendants. As a result, we have provided a dataset of three source videos designed and acquired for the purposes of the experiment. In addition, we have made them publicly available with the associated scores of the questionnaires and head-tracking of the participants. 

We can conclude that video quality assessment can be adapted to conditions imposed by socioemotional feature methodologies, such as contents of longer duration where the scene background is mainly static. This is an important contribution to the state of the art, since it shows that methodologies can be designed to simultaneously evaluate technical features and socioemotional features that go one step further. Thus, it allows this type of experiment in more realistic environments with final VR applications.

The prototype evaluated for VR communications provides high scores in terms of social and spatial presence. Significant differences in the sense of social presence have been obtained between sequences. Then, we can assure that social presence is highly influenced by the acquisition perspective, narrative, and non-verbal behaviour of the participants on the provider side, enriching the effectiveness of the conversation.

We have designed a questionnaire to evaluate attitude among participants. We have found significant differences between the experimental conditions and we can confirm that a positive impact has been achieved in all participants.

Finally, we cannot assure that the interactive element, the proper hands of the participants and a whiteboard with a whiteboard marker to take notes, significantly influences attention and spatial and social presence.

\ifCLASSOPTIONcompsoc
  \section*{Acknowledgments}
 This work has been partially supported by project PID2020-115132RB (SARAOS) funded by MCIN/AEI/10.13039/501100011033 of the Spanish Government and by project IDI-20200225 (TARDIS) funded by the Spanish Administration Agency CDTI. J. Gutiérrez was supported by a Juan de la Cierva fellowship (IJC2018-037816) of the Ministerio de Ciencia, Innovación y Universidades of the Spanish Government.  

\else
  \section*{Acknowledgment}
 
\fi


\ifCLASSOPTIONcaptionsoff
  \newpage
\fi



%

\bibliographystyle{IEEEtran}
\bibliography{biblio}

%

\begin{IEEEbiography}[{\includegraphics[width=1in,height=1.25in,clip,keepaspectratio]{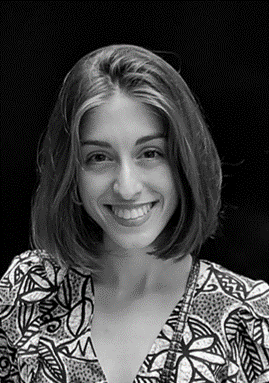}}]{Marta Orduna} received the Bachelor of Engineering in Telecommunication Technologies and Services in 2016 and the Master in Telecommunication Engineering (accredited by ABET) in 2018 (Master Graduation Award), both from the Universidad Polit\'ecnica de Madrid (UPM), Madrid, Spain. She has been a member of the Grupo de Tratamiento de Im\'agenes (Image Processing Group) of the UPM since 2016, where she has been actively involved in several research projects. Her current research is in the area of virtual reality, video encoding and streaming, and quality of experience.
\end{IEEEbiography}

\begin{IEEEbiography}[{\includegraphics[width=1in,height=1.25in,clip,keepaspectratio]{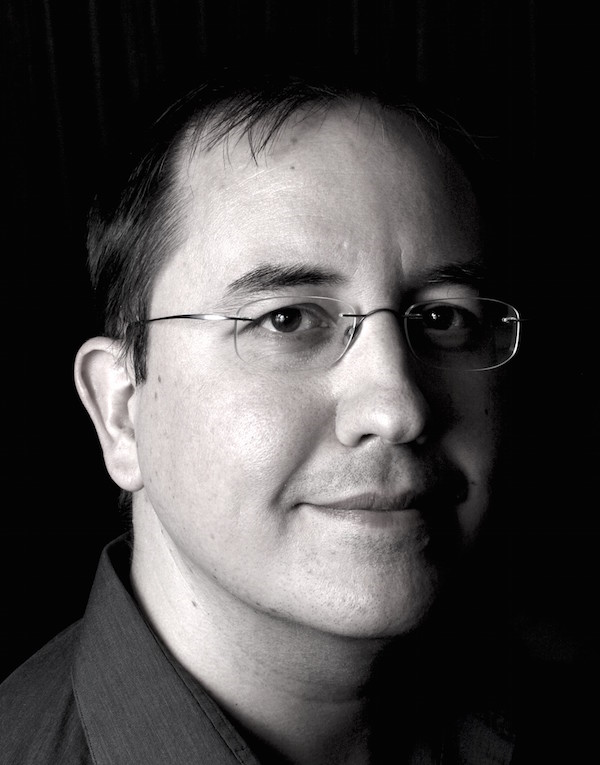}}]{Pablo P\'erez} received the Telecommunication Engineering degree (integrated BSc-MS) in 2004 and the Ph.D. degree in Telecommunication Engineering in 2013 (Doctoral Graduation Award), both from Universidad Polit\'ecnica de Madrid (UPM), Madrid, Spain. From 2004 to 2006 he was a Research Engineer in the Digital Platforms Television in Telef\'onica I+D and, from 2006 to 2017, he has worked in the R\&D department of the video business unit in Alcatel-Lucent (later acquired by Nokia), serving as technical lead of several video delivery products. Since 2017, he is Senior Researcher in the Distributed Reality Solutions department at Nokia Bell Labs. His research interests include multimedia quality of experience, video transport networks, and immersive communications.
\end{IEEEbiography}

\begin{IEEEbiography}[{\includegraphics[width=1in,height=1.25in,clip,keepaspectratio]{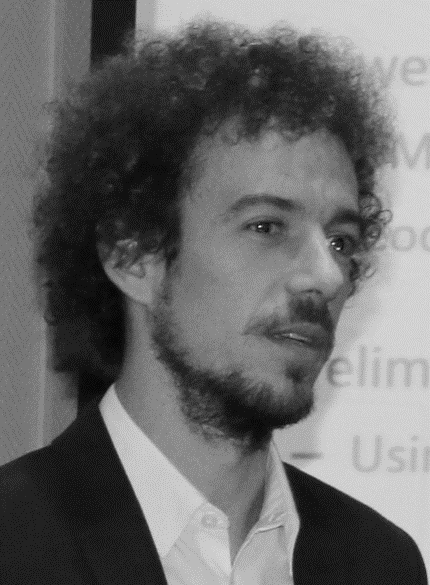}}]{Jesús Gutiérrez} is a research fellow (Juan de la Cierva) at the Image Processing Group (GTI) of the Universidad Politécnica de Madrid (UPM), Spain. He received the Telecommunication Engineering degree (five-year engineering program) from the Universidad Politécnica de Valencia (Spain) in 2008, the master’s degree in Communications Technologies and Systems (two-year M.S. program) in 2011, and the Ph.D. degree in Telecommunication in 2016, both from the UPM. From 2016 to 2019 he was a post-doctoral researcher (firstly, Marie Curie Fellow within the PROVISION ITN, and then, Marie Curie PRESTIGE Fellow) at the Image, Perception and Interaction (IPI) group of the Laboratoire des Sciences du Numérique de Nantes (LS2N) of the Université de Nantes (France). His research interests are in the area of image and video processing, evaluation of user quality of experience, immersive media technologies, and visual attention and human perception.
\end{IEEEbiography}

\begin{IEEEbiography}[{\includegraphics[width=1in,height=1.25in,clip,keepaspectratio]{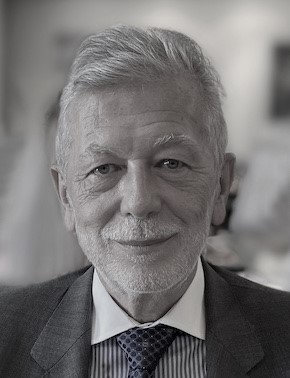}}]{Narciso Garc\'ia} received the Ingeniero de Telecomunicaci\'on degree (five years engineering program) in 1976 (Spanish National Graduation Award) and the Doctor Ingeniero de Telecomunicaci\'on degree (PhD in Communications) in 1983 (Doctoral Graduation Award), both from the Universidad Polit\'ecnica de Madrid (UPM), Madrid, Spain. Since 1977, he has been a member of the faculty of the UPM, where he is currently a Professor of Signal Theory and Communications. He leads the Grupo de Tratamiento de Im\'agenes (Image Processing Group), UPM. He has been actively involved in Spanish and European research projects, also serving as an evaluator, a reviewer, an auditor, and an observer of several research and development programs of the European Union. He was a Co-Writer of the EBU proposal, base of the ITU standard for digital transmission of TV at 34--45 Mb/s (ITU-T J.81). He was an Area Coordinator of the Spanish Evaluation Agency (ANEP) from 1990 to 1992 and he was the General Coordinator of the Spanish Commission for the Evaluation of the Research Activity (CNEAI) from 2011 to 2014. He has been the Vice-Rector for International Relations of the Universidad Polit\'ecnica de Madrid from 2014 to 2016. He was a recipient of the Junior and Senior Research Awards of the  Universidad Polit\'ecnica de Madrid in 1987 and 1994, respectively. His current research interests include digital video compression, computer vision, and quality of experience. 
\end{IEEEbiography}




\end{document}